**Theses by publication in Spain: output, authorship and integrity concerns**

Cabezas-Clavijo, Álvaro[1]; Barriga, Nazaret[2]; Repiso, Rafael[3]

[1]Universidad Internacional de La Rioja (UNIR)

[2]UNIE Universidad

[3]Universidad de Granada

This manuscript is a preprint and has not undergone peer review. Comments, suggestions, and constructive feedback are welcome and appreciated to help improve the quality and clarity of the final version.

**Abstract**

Theses by publication are accepted in most Spanish universities, yet there is little empirical evidence on the scientific output that supports them or on the role doctoral candidates play in that output, and it is not even possible to determine what share of all theses defended in the country follow this format. To address this gap, we analysed a sample of 338 theses by publication defended in Spain in 2025 and 2026, identified in the Teseo database and downloaded in full text. On average, each thesis included 4.0 works, of which 3.0 were articles indexed in Scopus, and 61.1% of these articles appeared in first-quartile SJR journals. The doctoral candidate was first author in 81.1% of the articles and held an intermediate position in 12.1%, while 17 candidates (5.4%) were not first or sole author of any article in their compendium. As for publication venues, MDPI was the main publisher, with 23.2% of the articles, and 40.6% of the candidates published at least one work from their compendium with it. Overall, 45.4% of the candidates used so-called *grey* publishers, such as MDPI or Frontiers, at least once, and for almost 10% of them these publishers were the only outlet for the compendium. We also detected other infrequent but questionable practices, such as the inclusion of works published in journals delisted from Scopus or in special issues guest-edited by the candidates themselves, works still under review, and articles included in two different theses. Taken together, these results point to the need to clarify regulations on theses by publication, in particular regarding the candidate's authorship of the works linked to the thesis and the quality criteria required of publication venues. Theses by publication should also be identified in a standardised way in official registers, so that they can be retrieved and monitored.

## Introduction

A thesis by publication is a form of doctoral thesis whose body consists, wholly or mainly, of works already published or accepted in peer-reviewed journals or books (UK Council for Graduate Education [UKCGE], 1996), brought together around a single line of research so that they form a coherent and relevant contribution to knowledge rather than a mere collection of papers (Ortega Martínez, 2014). Because it departs from the traditional monograph thesis, it has been regarded as an innovation in doctoral education (de Granda-Orive & Villena-Garrido, 2018; Ortega Martínez, 2014), and it has become a widely used route because of the advantages it offers to doctoral candidates and supervisors. Despite its current importance, this format has received little attention in the academic literature on Spain. According to Teseo[1], the main source of information on theses in the country, around 11,000 doctoral theses were defended each year between 2021 and 2024, but it is not known what share of them were theses by publication.

According to data for 2022[2], the most recent available, 84.7% of candidates received the highest grade (*sobresaliente cum laude*). There are no data on theses rated as "fail" at the defence, since no official record of such cases exists, and it is almost unheard of for a candidate to fail the doctoral examination, because quality assurance for theses operates beforehand rather than at the defence. In addition, examination committees are often proposed by the candidate and the supervisors, with varying degrees of intervention, on the basis of friendship (Repiso, 2013) or at least cordial relations, which can lead to complacency in assessment. As a result, legitimate questions have been raised about the quality of theses defended in Spain. Jiménez-Contreras et al. (2014) note that academia has avoided confronting the reality of theses approved as a mere formality and, ultimately, of poor theses. It is therefore not surprising that the thesis by publication is seen as an additional source of legitimacy, since it gives the thesis a double certification: the validation of the examination committee, often lenient, is complemented by that of the academic community, which certifies the candidate's research through peer review.

---

[1] https://aplicaciones.ciencia.gob.es/teseo/#/home
[2] https://www.ciencia.gob.es/InfoGeneralPortal/documento/2f2814bc-4753-4d4e-a69f-67785415850a

**Evolution of doctoral studies**

There is no consensus on who was the first doctor in history or which institution awarded the first degree. Degrees emerged gradually in the first European *studia* in the twelfth and thirteenth centuries, and both Bologna and Paris claim precedence (Noble, 1994; Verger, 1992). There is agreement, however, that the doctorate was created by the medieval European university (Archbald, 2011) and that these two models established the hierarchy of degrees (bachelor, licentiate and doctor, or master in the Parisian case), which other universities inherited (Cobban, 1975). Originally, the degree did not certify research but recognised the scholarly authority of an academic. It was the highest distinction awarded by the university corporation and carried the *licentia ubique docendi*, the right to teach at any university in Christendom, which at the time made up the whole university system (Verger, 1992), although many universities later required doctors to pass formal examinations (Rashdall, 1936). In the medieval university, the doctoral degree was obtained through a private examination and a public defence before the faculty, in which the candidate demonstrated mastery of the discipline and fitness to teach it (Noble, 1994; Taylor, 2023). In other words, for centuries there were doctorates without doctoral theses.

The doctorate as a credential certifying research ability came much later and originated in the Prussian university (Cuthbert & Molla, 2015). Specifically, the University of Berlin, founded in 1810 at the initiative of Wilhelm von Humboldt, introduced a doctorate awarded for an original contribution to knowledge (Taylor, 2023). Outside Germany, most European universities remained focused on transmitting received knowledge, and doctoral training consisted largely of lectures, reading, recitation and the reproduction of other people's writings (Archbald, 2011).

Spain belongs to this second group. Alejo Montes (2007) notes that, in the Salamanca model of the Ancien Régime, the examination of knowledge belonged to the licentiate, while the doctorate was "mere ceremony and festivity". The thesis as a document appeared with the Pidal Plan of 1845, still as a printed speech that the new doctor read and defended before the faculty at the investiture ceremony, once the examinations had been passed (Repiso, 2013). The Moyano Law of 1857, centralist and French-inspired, reserved the award of the doctoral degree to the Central University, and the Royal Decree of 22 May 1859, which developed the Moyano Law in detail, turned that speech into the very object of assessment, although the topic was chosen from a list prepared by the faculty (Repiso, 2013).

The change was formal rather than scientific: these early theses are considered to be of low quality (Petit, 1997), and the requirement of an original research work on a freely chosen topic did not arrive until the Regulation on Examinations and Degrees of the Central University in 1900 (Repiso, 2013). To give an idea of what this meant, the thesis of Nobel laureate Santiago Ramón y Cajal (1877), is essentially discursive in tone, handwritten and 45 pages long in A5 format.

**Theses by publication: background, advantages and drawbacks compared with the monograph thesis**

The informal precedent of the thesis by publication is the doctorate awarded to Ludwig Wittgenstein in 1929 for the *Tractatus Logico-Philosophicus*, published eight years earlier (Davies & Rolfe, 2009, cited in Ortega Martínez, 2014). Institutionally, however, the thesis by publication has no single origin. In continental Europe, submitting published works as the basis for a doctorate was already established practice in countries such as Belgium, the Netherlands and Sweden, although in none of them was it the only route to the degree (UKCGE, 1996; Powell, 2004). In the United Kingdom, its institutionalisation has a precise date: in the early 1960s the University of Cambridge set up a committee on the award of the doctorate, and in 1966 it decided to introduce the PhD by published work, initially open to former students, with the first thesis submitted in social anthropology (UKCGE, 1996; Powell, 2004). During its first thirty years, the British route was seen as little used and was reserved mainly for the institution's own staff and, to a lesser extent, its graduates (UKCGE, 1996); it was also governed by different regulations from the conventional route. Its expansion is recent: the number of British universities offering it rose from 37 of 73 (51%) in 1996 to 64 of 80 (80%) in 2004, and the number of degrees awarded over a two-year period rose from 52 to 116 (Powell, 2004).

The thesis by publication has several advantages over the monograph thesis. The main ones are the following:

1. It generates greater research output for the candidate and the university, in contrast to the limited impact of the traditional thesis (Larivière et al., 2008), which benefits the university in rankings and in project reporting and makes the candidate's CV more competitive. In a study of German social science theses, theses by publication led to three times as many publications as monographs, and these publications also received three times as many citations (Asanov et al., 2024).
2. It adds an extra layer of control by involving reviewers external to the thesis, which can give these theses additional legitimacy. Thesis examiners themselves acknowledge that prior peer review gives them an additional guarantee of the quality of the work (Robins & Kanowski, 2008; Sharmini, 2016), although anonymous review does not by itself guarantee the quality of the thesis (Ortega Martínez, 2014).
3. It increases the likelihood that the results of the thesis will be published since, as Fridlund (2010) notes, a new professional stage usually begins after the defence, which makes it harder to turn the thesis into publications.
4. It allows the scientific community to access these results and assess their quality, as reviewers and readers (Merga et al., 2020), whereas traditional monographs have historically been hard to access.

5. It prevents data from becoming obsolete, a problem that often jeopardises the publication of thesis-based work in fields where knowledge evolves quickly (Merga, 2015; Robins & Kanowski, 2008).
6. It places the candidate in the usual space for research communication, the scientific journal, so that when candidates enter the world of scholarly communication (Craig et al., 2026) they are not only consumers but also producers.
7. It allows a large project to be divided into manageable, well-defined parts (Robins & Kanowski, 2008), which makes it easier to plan the doctorate and manage time (Merga et al., 2020), a valuable skill for the researcher's later career.
8. It often brings the candidate into large research teams (Fridlund, 2010), sometimes as a result of research stays. This provides a basic structural dimension, since modern science in most disciplines is carried out collectively by multidisciplinary, multi-institution teams, whereas the monograph thesis is usually confined to the relationship between master and disciple.

However, this format also has several limitations that should be considered.

1. The thesis by publication depends on editorial timelines (Merga et al., 2020). Peer review, acceptance and publication do not depend on the candidate or the institution, and they can extend the doctorate unpredictably. This is particularly problematic in programmes with strict time limits, where it may even lead to the candidate's dismissal from the programme (Mertkan et al., 2022) and prevent them from obtaining the degree.

2. Conflicts may arise from the transfer of exploitation rights to publishers (Craig et al., 2026; Ortega Martínez, 2014). Depending on each journal's policy, these conflicts may restrict or condition the full reproduction of the articles in the final thesis and its subsequent open deposit.
3. In some fields, notably the humanities, where the format is less established (Sharmini, 2016), it may deprive the candidate of a central formative experience: writing an extended monograph. In these disciplines, the monograph remains the main vehicle for scholarly communication, and writing one requires knowledge and skills different from those needed for an article, such as sustaining an argument over tens or hundreds of pages, building a broader textual architecture and integrating data and information on a large scale. Writing a book is also a way of thinking about and structuring one's own research (Crossick, 2015), so the absence of this experience may be a significant gap in the training of future researchers, especially in the humanities.
4. Another problem is that journal publication times need not match the length of the doctorate, and this pressure can encourage poor practices by both candidates and supervisors. These include choosing unsuitable or dubious journals, with more concern for the speed and ease of the editorial process than for their suitability, or even the unjustified fragmentation of results into minimal publications (*salami slicing*) (de Granda-Orive & Villena-Garrido, 2018).

5. Finally, the greatest threat to the credibility of the system is the uncertainty about the candidate's real contribution to co-authored works (Elmgren et al., 2024; Goddiksen et al., 2023; Sharmini, 2016). By the end of their training, doctors should master every stage of a scientific work, from formulating the problem to writing up and defending the results. However, the collective nature of publications in some fields can mask cases of substitution or undue attribution of authorship, and research groups may even jointly sustain articles that support theses whose intellectual authorship does not, in substance, belong to the candidate.

## Theses by publication in Spain

It is not possible to determine when the thesis by publication first appeared in Spain, since no national regulation has ever governed it and it has spread under university autonomy, on the understanding that the originality required of a thesis does not mean it must be unpublished (Ortega Martínez, 2014). The earliest documented reference is the University of Lleida, which approved its criteria in February 2002, and by 2014, 59 of the 80 Spanish universities already accepted it (Ortega Martínez, 2014).

Since then, the regulatory framework has not changed substantially. Royal Decree 99/2011 of 28 January defines the doctoral thesis as an original research work produced by the candidate in any field of knowledge (art. 13.1), without specifying its format, and neither this decree nor its later amendments (the most recent through Royal Decree 576/2023 of 4 July) mention theses by publication. Their regulation is therefore left to universities and their doctoral schools, which generally accept them (Jiménez-Ramírez, 2017), with the resulting wide variation in requirements.

Ortega Martínez's (2014) analysis of the regulations of 56 universities illustrates this heterogeneity. Only 15 had specific regulations for this type of thesis, usually one or two pages long, while in the rest the rules were limited to a few lines within the general doctoral regulations. The most common minimum number of publications was three (55.4% of universities), although 26.8% did not specify one, and 37.5% set no criterion on the standing of the journals; when they did, the most frequent reference was the Journal Citation Reports. Co-authorship was accepted in 83.9% of universities, but almost half (48.2%) did not specify the position the candidate should hold in the publications, and fewer than 20% stated that the candidate should preferably be first author. Ortega Martínez (2014) concluded that the thesis by publication was an important innovation, but one accompanied by "a great deal of improvisation".

A more recent analysis of theses by publication in six countries, including Spain, based on 50 regulations (Mason et al., 2025), finds that the most common model requires the contributions to have been published during the doctoral period rather than before enrolment in the programme, sets a minimum number of publications, usually three, and shows a clear preference for peer-reviewed articles. It also expects the candidate to be the lead author of all or most of the works in the compendium.

This issue has also been examined in specific fields such as surgery. Serra-Aracil et al. (2022) analysed the requirements for theses by publication in 29 Spanish university departments and found notable differences between them. Most (83%) required a minimum of two articles, and 41% raised that threshold to three or more. As for authorship, in almost half of the cases the candidate had to appear in a leading position in at least two of the articles. Finally, 48% of departments set no requirement on the JCR quartile of the journals.

The institutional framework must also be considered. Doctoral programmes at Spanish universities are verified and accredited by the National Agency for Quality Assessment and Accreditation (ANECA), which, among other criteria, uses the scientific contributions derived from doctoral theses as an indicator of quality (Jiménez-Contreras et al., 2014). There are therefore clear institutional incentives to promote this format, in which publication is part of the requirement for the defence and which increases the output of doctoral programmes.

In any case, the debate on the safeguards of this format remains open. De Granda-Orive and Villena-Garrido (2018) warned that the thesis by publication requires the candidate's contribution as an author to be clearly established and demands full commitment from supervisors, and they also pointed to the risk of fragmenting results under pressure to publish. This emphasis on productivity can be experienced in very different ways: for some candidates, publishing during the thesis is a stimulus that supports their professional development and their entry into an academic career (Wilkins et al., 2021), while for others it adds pressure at an already demanding stage (Horta & Li, 2023).

Although theses by publication are now accepted in most Spanish universities, there are no data on how many theses are defended through this route or on the characteristics of the output that supports them. Teseo, the official register of doctoral theses in Spain, has no field identifying the format, and existing studies have focused on analysing regulations, without empirically examining productivity, the candidate's role in the publications, their academic impact or the dissemination channels used. This study aims to fill that gap.

**Objectives**

The general objective of this study is to characterise, from a bibliometric perspective, the doctoral theses by publication defended in Spain in 2025 and 2026 and the scientific output on which they are based. This objective is broken down into the following specific objectives (SO):

SO1. To identify the theses by publication registered in Teseo in the period analysed and describe their distribution by field of study and by the candidate's sex.

SO2. To quantify the productivity of these theses, comparing the number of works declared in the compendium with the articles retrieved from Scopus, and to analyse differences between fields.

SO3. To analyse the candidate's position in the byline of the articles linked to the thesis, as a proxy for their contribution to the research.

SO4. To assess the visibility and impact of this output through the citations received and the SJR quartile of the publishing journals.

SO5. To identify the journals and publishers through which the output of theses by publication is disseminated, and their degree of concentration.

SO6. To examine whether there are differences by sex in the above indicators.

**Methodology**

The sample of theses by publication was drawn from Teseo[3], a database set up by the Spanish Ministry of Education and Science that has recorded the doctoral theses defended in Spain since 1976 and serves as the centralised official register. Under current Spanish doctoral regulations (Royal Decree 99/2011 of 28 January, regulating official doctoral studies, art. 14.5), universities are required to submit this information to Teseo. However, because universities organise this procedure in different ways, Teseo's coverage differs slightly from the official data published by the Ministry (by around 2% in recent years with comparable data[4]), and there may also be delays in uploading information.

We searched this database by content, that is, in the title and abstract fields of the doctoral thesis, using the terms "compendio" and "*compendium*". We also tested the English terms most frequently used in the international literature ("thesis by publication", "PhD by publication", "PhD by published work", "compilation thesis" and "article-based thesis"), but they did not retrieve any additional theses. The search was carried out on 26 May 2026, repeated on 10 July and updated again on

[3] https://aplicaciones.ciencia.gob.es/teseo/#/home

[4] https://www.ciencia.gob.es/InfoGeneralPortal/documento/2f2814bc-4753-4d4e-a69f-67785415850a

11 October of the same year, and covered theses defended in 2025 and 2026, in the latter case up to the date of data collection. In total, 434 doctoral theses were identified as theses by publication, which represents 3.3% of the 12,999 theses recorded in Teseo for 2025-2026 at the date of data collection.

It should be noted that this is not the total number of theses by publication defended in Spain, but only those that declared this format in the thesis abstract. It is currently not possible to determine the total number of theses by publication defended in Spain, since neither Teseo nor other thesis databases such as Dialnet have a classification or filter to identify them. The sample is therefore a non-probabilistic purposive sample, and the results should be interpreted as describing the set analysed rather than as representative of all theses by publication.

Once the list had been compiled, we downloaded the full text of the theses, either from Teseo or from other sources such as institutional repositories. Under the official regulations (Royal Decree 99/2011 of 28 January, regulating official doctoral studies, art. 14.5 and 14.6), theses must be deposited in open electronic format in an institutional repository, except in exceptional circumstances such as the participation of companies in the doctoral programme, confidentiality agreements with companies or the possibility of generating patents.

Through this process, we were able to download the full text of 346 theses (79.7% of the theses recorded in Teseo). In the remaining cases (88 theses), the download was not possible because the document was under embargo and not available for consultation, because of technical or access errors, or because it was not yet accessible through an institutional repository.

Each of the 346 downloaded theses was examined manually to determine the total number of publications that each candidate included in the compendium. In this process, we found that eight theses contained the word "compendio" in the abstract but did not belong to this format, so they were excluded, and the final sample consisted of the 338 theses by publication that could be downloaded.

Once the number of publications listed in each thesis had been recorded, we searched for and downloaded them from Scopus, a multidisciplinary abstract and citation database that indexes journals, conference proceedings and books selected by an independent committee of experts, and which has become one of the main data sources for research evaluation, science policy analysis and university rankings (Baas et al., 2020). We recorded the number of documents that could be identified in Scopus and downloaded a CSV file with the scientific output of each candidate included in the thesis document.

In addition, for each of these works we recorded the candidate's byline position (first author, last author, sole author or intermediate author), a task performed with Claude (Sonnet 5 model) and checked by the authors. We also recorded, for each work, the number of citations received at the last data download from Scopus (11 and 12 October 2026), and the journal quartile in the Scimago Journal & Country Rank (SJR)[5] according to SJR 2025 data, taking the journal's best quartile across its subject categories. The "No quartile" category includes, on the one hand, sources included in SJR without an assigned quartile and, on the other, sources not included in SJR 2025, such as conference proceedings, book chapters and delisted journals.

[5] https://www.scimagojr.com/

For the analysis by publisher, the publisher names recorded in Scopus were grouped under the publishing group to which they belong, so that the same publisher would not appear split across several entries. For example, the different trading names of Elsevier (Elsevier Ltd, Elsevier B.V., Elsevier Inc., etc.) were merged into "Elsevier", while imprints such as BioMed Central or Nature Publishing Group were included in "Springer Nature". This normalisation was also performed with Claude (Sonnet 5 model).

For the classification by field, we used Teseo's own taxonomy, which identifies ten fields of study and assigns each thesis to one of them[6]. The candidate's sex was inferred from their first name using the Python library *gender_guesser*, run through Claude (Sonnet 5 model), and complemented by a table of exceptions that was checked manually for names that were infrequent in the sample.

Finally, differences between groups were tested with the Mann-Whitney U test for continuous variables, since they were not normally distributed, and with Pearson's $\chi^2$ test for comparisons of proportions. The significance level was set at 0.05. Data processing and statistical analyses were carried out in Python (pandas and SciPy libraries), and the figures were produced in R (ggplot2 package).

## Results

[6] The fields of study and abbreviations used are: Health and welfare (HEA); Education (EDU); Social sciences, journalism and information (SOC); Natural sciences, mathematics and statistics (NAT); Engineering, manufacturing and construction (ENG); Arts and humanities (ART); Information and Communication Technologies (ICT); Business, administration and law (BUS); Agriculture, forestry, fisheries and veterinary (AGR); and Services (SER).

The 338 theses yielded a total output of 1350 works, of which 1002 (74.2%) could be retrieved from Scopus. Mean productivity per thesis was therefore 4.0 works (3.0 Scopus articles), slightly higher among men than among women (3.1 vs 2.8 Scopus articles), although the difference was not statistically significant (Mann-Whitney U, $p = 0.127$). This output received a mean of 9.9 citations per article (median 4), significantly higher in theses defended by men (11.1 vs 8.2; Mann-Whitney U, $p = 0.004$) (Table 1).

Table 1: General indicators of the sample

| Indicator | ALL | MEN | WOMEN |
|---|---|---|---|
| No. of theses analysed | 338 | 195 | 143 |
| Works in compendium | 1350 | 793 | 557 |
| Articles in Scopus | 1002 | 600 | 402 |
| % articles in Scopus | 74.2 | 75.7 | 72.2 |
| Mean works per thesis | 4.0 | 4.1 | 3.9 |
| Mean Scopus articles per thesis | 3.0 | 3.1 | 2.8 |
| Median Scopus articles per thesis | 3 | 3 | 3 |
| Total citations | 9967 | 6663 | 3304 |
| Mean citations per article | 9.9 | 11.1 | 8.2 |
| Median citations per article | 4 | 5 | 4 |

The analysis by field of study (Table 2), based on the Teseo classification, shows that Health and welfare (108) is the field with the most theses by publication, followed by Education (46) and Social sciences, journalism and information (45). Productivity per thesis is highest in Natural sciences, mathematics and statistics, with a mean of 4.1 Scopus articles, followed by Engineering, manufacturing and construction, with 3.8 Scopus articles per thesis.

The largest gaps between the output declared in the thesis and the articles retrieved from Scopus are found in Services (from 5 works to a single article, although only two theses were analysed), Arts and humanities (from 3.9 works declared in the compendium to 1.5), Education (a difference of 2.0 works), Social sciences, journalism and information (1.5) and Business, administration and law (1.4).

Citation rates are highest in Engineering, manufacturing and construction (15.0 citations per article) and Information and Communication Technologies (13.7), and lowest in Services (0.5 citations), Arts and humanities (4.4) and Social sciences, journalism and information (6.8). These marked differences reflect, as expected, the different citation practices of each field.

Table 2: General indicators by field of study

| **Indicator** | HEA | EDU | SOC | ENG | NAT | ART | ICT | BUS | AGR | SER |
|---|---|---|---|---|---|---|---|---|---|---|
| No. of theses analysed | 108 | 46 | 45 | 36 | 35 | 25 | 15 | 18 | 8 | 2 |
| Works in compendium | 413 | 198 | 187 | 150 | 149 | 97 | 53 | 67 | 26 | 10 |
| Articles in Scopus | 347 | 107 | 118 | 136 | 142 | 37 | 52 | 41 | 20 | 2 |
| % articles in Scopus | 84.0 | 54.0 | 63.1 | 90.7 | 95.3 | 38.1 | 98.1 | 61.2 | 76.9 | 20.0 |
| Mean works per thesis | 3.8 | 4.3 | 4.2 | 4.2 | 4.3 | 3.9 | 3.5 | 3.7 | 3.3 | 5.0 |
| Mean Scopus articles per thesis | 3.2 | 2.3 | 2.6 | 3.8 | 4.1 | 1.5 | 3.5 | 2.3 | 2.5 | 1.0 |
| Median Scopus articles per thesis | 3 | 3 | 3 | 3 | 4 | 1 | 3 | 2.5 | 3 | 1 |
| Total citations | 3250 | 812 | 808 | 2035 | 1660 | 162 | 711 | 362 | 166 | 1 |
| Mean citations per article | 9.4 | 7.6 | 6.8 | 15.0 | 11.7 | 4.4 | 13.7 | 8.8 | 8.3 | 0.5 |
| Median citations per article | 4 | 3 | 2 | 9 | 5.5 | 1 | 8 | 4 | 4 | 0.5 |

The histogram of Scopus productivity (Figure 1) shows that the modal value per thesis is 3 articles (41.1% of cases), followed by 4 articles (16.9%) and 2 articles (13.6%). In 7.4% of cases, no Scopus article linked to the thesis by publication was found.

Figure 1: Histogram of Scopus article productivity per thesis

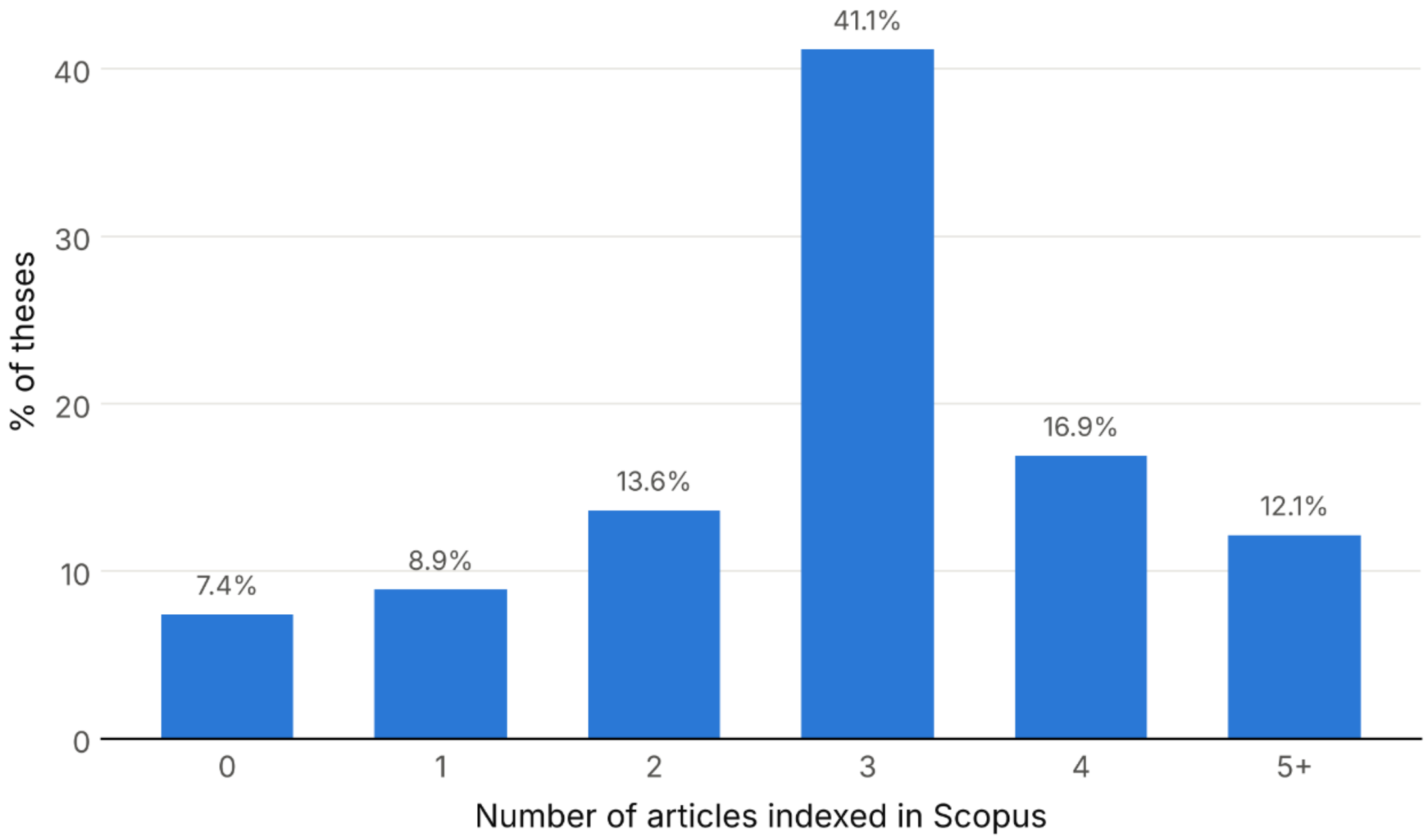

By field of study, the field with the highest proportion of theses with no Scopus articles or only one is Arts and humanities, where 60% of theses include one or no Scopus article, followed by Education (28.3%) and, very closely, by Business, administration and law (27.8%). In contrast, in both Natural sciences, mathematics and statistics and Information and Communication Technologies, all theses include at least 2 Scopus articles. The highest percentage of theses with 4 or more works is found in Natural sciences, mathematics and statistics (51.4%) and Engineering, manufacturing and construction (41.7%), followed by Information and Communication Technologies (33.3%) (Figure 2).

Figure 2: Histogram of Scopus article productivity per thesis by field of study[7]

[7] Agriculture, forestry, fisheries and veterinary, and Services are excluded from this figure because of the small number of theses in these fields.

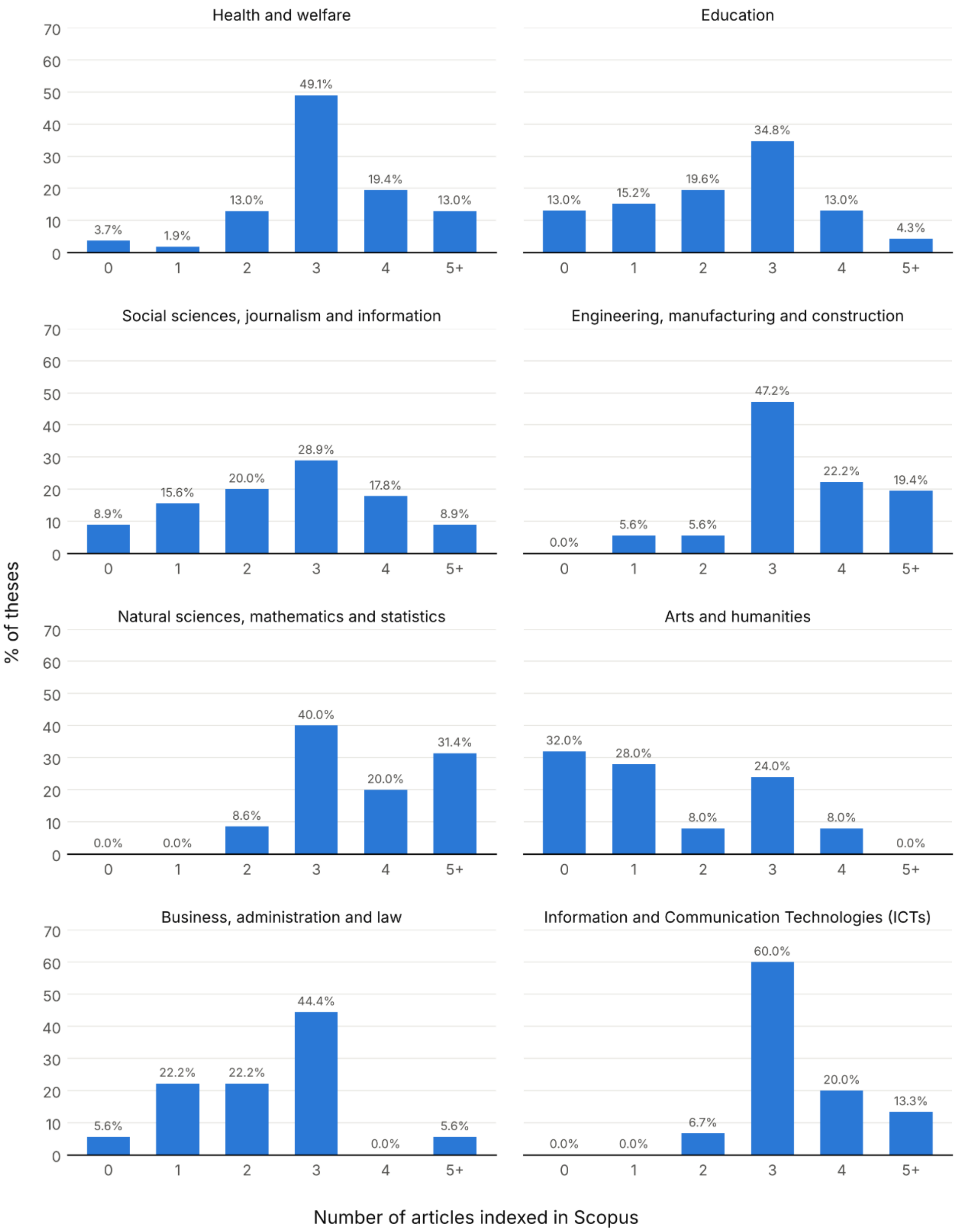

As for the candidate's authorship position in the works linked to the thesis (Table 3), the candidate is first author in most cases (81.1%), a position that signals a leading role in the research. However, it is striking that in 12.1% of the works the candidate holds an intermediate position, a figure that rises to 14.0% for men and falls to 9.2% for women, with the difference being statistically significant ($\chi^2 = 4.77$; $p = 0.029$).

Table 3: Number and % of authorships by the candidate's position in the byline

| | ALL | | MEN | | WOMEN | |
|---|---|---|---|---|---|---|
| **Authorship position** | **N** | **%** | **N** | **%** | **N** | **%** |
| First | 813 | 81.1 | 477 | 79.5 | 336 | 83.6 |
| Intermediate | 121 | 12.1 | 84 | 14.0 | 37 | 9.2 |
| Last | 24 | 2.4 | 14 | 2.3 | 10 | 2.5 |
| Sole | 44 | 4.4 | 25 | 4.2 | 19 | 4.7 |
| **Total** | **1002** | **100** | **600** | **100** | **402** | **100** |

At the level of fields, intermediate authorship by the candidate is more common in Information and Communication Technologies, where the candidate holds an intermediate position in 21.2% of the publications linked to the compendium, followed by Agriculture, forestry, fisheries and veterinary (20.0%) and Health and welfare (16.4%). In contrast, no intermediate authorship by the candidate was found in three fields (Table 4).

Table 4: Percentage of authorships by the candidate's position in the byline and field of study

| **Authorship position** | HEA | EDU | SOC | ENG | NAT | ART | ICT | BUS | AGR | SER |
|---|---|---|---|---|---|---|---|---|---|---|
| First | 81.8 | 80.4 | 81.4 | 88.2 | 81.0 | 51.4 | 78.8 | 85.4 | 75.0 | 100 |
| Intermediate | 16.4 | 9.3 | 5.1 | 10.3 | 13.4 | 0 | 21.2 | 0 | 20.0 | 0 |
| Last | 1.2 | 5.6 | 1.7 | 1.5 | 4.2 | 0 | 0 | 7.3 | 5.0 | 0 |
| Sole | 0.6 | 4.7 | 11.9 | 0 | 1.4 | 48.6 | 0 | 7.3 | 0 | 0 |
| **N (articles)** | **347** | **107** | **118** | **136** | **142** | **37** | **52** | **41** | **20** | **2** |

The output linked to theses by publication appeared mainly in first-quartile journals (61.1%), followed at a considerable distance by Q2 journals (24.4%), according to SJR 2025, while slightly less than 15% appeared in third- or fourth-quartile journals or in sources with no quartile (Table 5). There were no significant differences by sex ($\chi^2$, p = 0.536), although the output of women candidates tends to appear slightly more often in Q1 journals (62.4% vs 60.2% for men), while men publish slightly more in Q2 (24.8% vs 23.6%).

Table 5: Number and % of articles by SJR 2025 quartile

| | ALL | | MEN | | WOMEN | |
|---|---|---|---|---|---|---|
| **Quartile** | **N** | **%** | **N** | **%** | **N** | **%** |
| Q1 | 612 | 61.1 | 361 | 60.2 | 251 | 62.4 |
| Q2 | 244 | 24.4 | 149 | 24.8 | 95 | 23.6 |
| Q3 | 87 | 8.7 | 51 | 8.5 | 36 | 9.0 |
| Q4 | 28 | 2.8 | 16 | 2.7 | 12 | 3.0 |
| No quartile | 31 | 3.1 | 23 | 3.8 | 8 | 2.0 |
| **Total** | **1002** | **100** | **600** | **100** | **402** | **100** |

By field of study, the fields with the highest share of Q1 output are Agriculture, forestry, fisheries and veterinary (80.0%, although with only 20 articles), Engineering, manufacturing and construction (70.6%) and Health and welfare (66.3%). In contrast, Services (no Q1 output, although with only two works), Information and Communication Technologies (40.4%) and Social sciences, journalism and information (49.2%) show the lowest output in these high-visibility journals (Table 6).

Table 6: Percentage of articles by SJR 2025 quartile and field of study

| **Indicator** | HEA | EDU | SOC | ENG | NAT | ART | ICT | BUS | AGR | SER |
|---|---|---|---|---|---|---|---|---|---|---|
| % Q1 | 66.3 | 53.3 | 49.2 | 70.6 | 63.4 | 56.8 | 40.4 | 56.1 | 80.0 | 0 |
| % Q2 | 21.9 | 23.4 | 27.1 | 18.4 | 31.0 | 29.7 | 42.3 | 19.5 | 5.0 | 0 |
| % Q3 | 9.5 | 13.1 | 12.7 | 2.9 | 4.9 | 10.8 | 5.8 | 14.6 | 5.0 | 0 |
| % Q4 | 2.3 | 3.7 | 7.6 | 1.5 | 0.7 | 0 | 0 | 4.9 | 5.0 | 50.0 |
| No quartile | 0 | 6.5 | 3.4 | 6.6 | 0 | 2.7 | 11.5 | 4.9 | 5.0 | 50.0 |
| **N (articles)** | **347** | **107** | **118** | **136** | **142** | **37** | **52** | **41** | **20** | **2** |

Table 7 shows the main journals, as well as the publishers most used for the works linked to the compendium. Journals are highly dispersed (639 different sources), with Applied Sciences (MDPI), Sustainability (MDPI) and Journal of Clinical Medicine (MDPI) showing the highest output. At the same time, output is highly concentrated in a small number of multinational publishing groups: 23.2% of the articles were published in MDPI journals, followed by 16.0% in Elsevier and 13.3% in Springer Nature, so that together these three publishers account for more than half (52.4%) of the candidates' output linked to their theses by publication.

Table 7: Main journals and publishers

| Journal | N | % | Publisher | N | % |
|---|---|---|---|---|---|
| Applied Sciences (Switzerland) | 17 | 1.7 | MDPI | 232 | 23.2 |
| Sustainability (Switzerland) | 16 | 1.6 | Elsevier | 160 | 16.0 |
| Journal of Clinical Medicine | 15 | 1.5 | Springer Nature | 133 | 13.3 |
| Healthcare (Switzerland) | 14 | 1.4 | Wiley | 47 | 4.7 |
| Journal of High Energy Physics | 14 | 1.4 | Frontiers | 34 | 3.4 |
| Frontiers in Psychology | 12 | 1.2 | Taylor & Francis | 33 | 3.3 |
| Sensors | 12 | 1.2 | IEEE | 23 | 2.3 |
| International Journal of Environmental Research and Public Health | 10 | 1.0 | SAGE | 19 | 1.9 |
| Bioengineering | 8 | 0.8 | Emerald Publishing | 11 | 1.1 |
| Nursing Reports | 8 | 0.8 | BMJ | 10 | 1.0 |
| BMJ Open | 8 | 0.8 | Ediciones Doyma | 9 | 0.9 |

MDPI's dominance as the main publisher rests above all on Health and welfare, where 30.3% of the output appears in its journals. It is also the leading publisher in Education, in Social sciences, journalism and information, and in Arts and humanities, although with shares below 18%, and it shares first place with Wiley in Business, administration and law (14.6%). Elsevier is the most used publisher in Engineering, manufacturing and construction (30.9%, very close to MDPI, with 30.1%), in Information and Communication Technologies (23.1%) and in Agriculture, forestry, fisheries and veterinary (35.0%), whereas Springer Nature is the most used in Natural sciences, mathematics and statistics (24.6%, closely followed by Elsevier, with 23.9%) (Figure 3).

Figure 3: Distribution of publishers by field of study

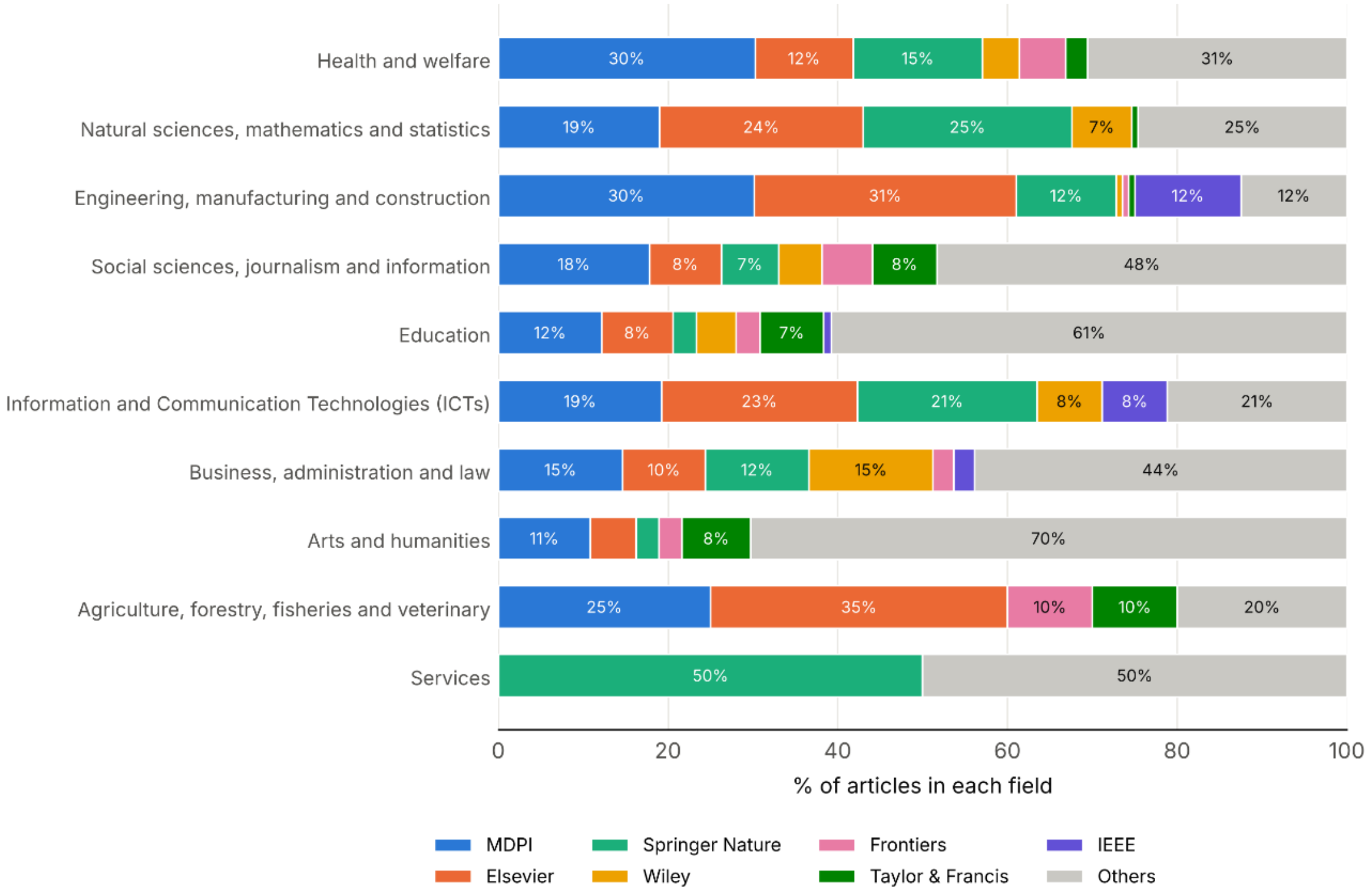

From the perspective of the candidates (Figure 4), MDPI again ranks first: 40.6% of the candidates with at least one Scopus article published at least one work from their compendium with this publisher, followed by Elsevier (32.9%) and Springer Nature (27.2%), and further behind by Wiley (13.4%), Frontiers and Taylor & Francis (8.6% each). Overall, 45.4% of the candidates (142 of 313) have at least one work with so-called *grey* publishers, such as MDPI or Frontiers (Nicholas et al., 2023), and for 9.6% (30 candidates) these two publishers are the only publication channel for their compendium.

Figure 4: Percentage of candidates with at least one article with the main publishers

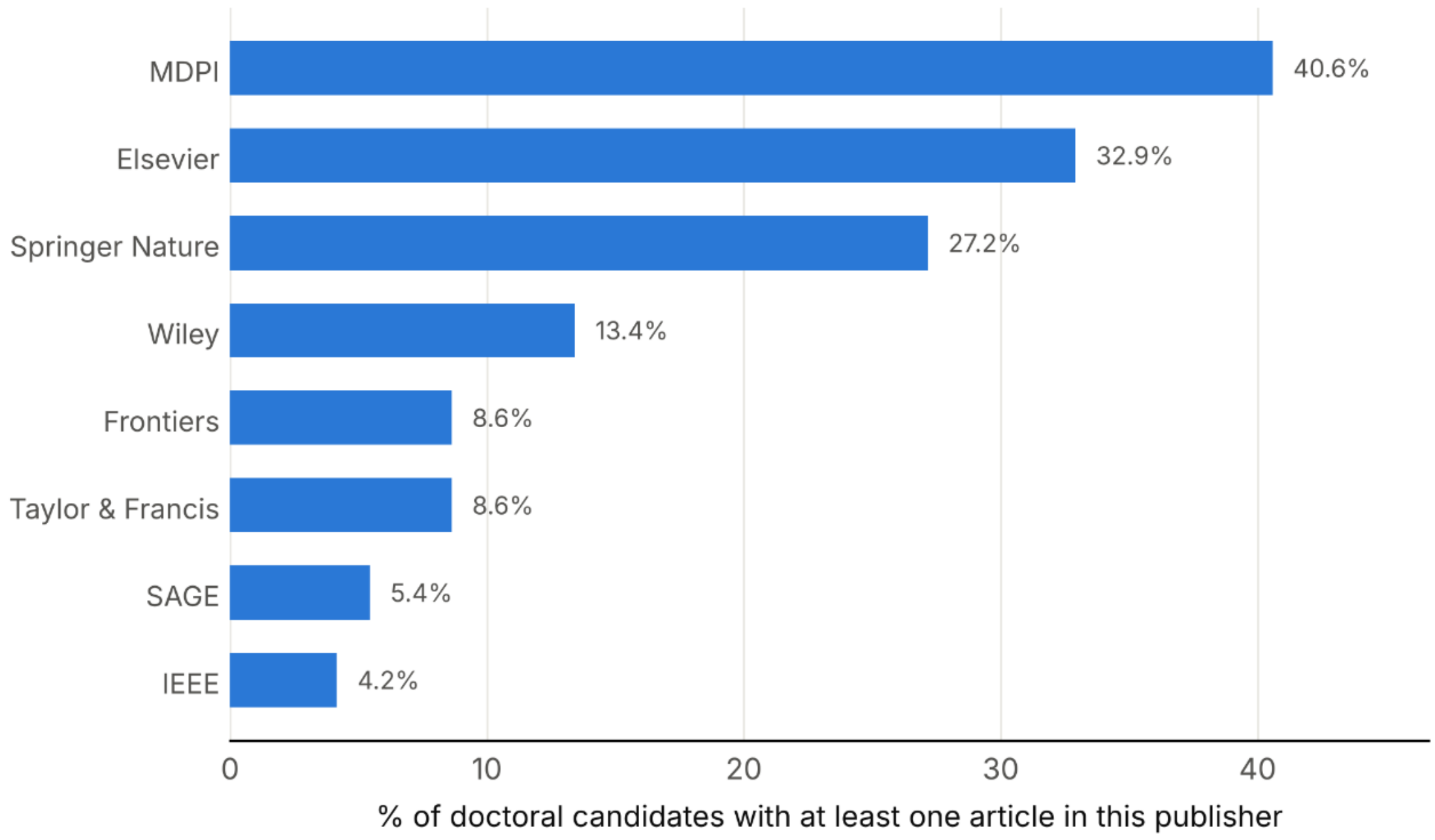

## Discussion and conclusions

This study offers the first national, multidisciplinary analysis of the productivity and bibliometric characteristics of theses by publication defended in Spain. Based on a sample of 338 doctoral theses from different fields, with a predominance of Health and welfare (32.0% of theses), our study has established the main characteristics of this form of doctoral research.

A thesis by publication defended in Spain includes on average 4 works, of which 3 are articles indexed in Scopus, figures consistent with the regulations governing theses by publication in Spain, which generally require at least three research works (Ortega Martínez, 2014). In these works, the candidate usually appears as the lead author (normally first author), and the works are published in high-visibility Scopus journals (mainly Q1 and Q2), mostly by large multinational publishers, especially MDPI and Elsevier.

There are few differences in productivity by sex (3.1 Scopus articles for men vs 2.8 for women), although the output of men candidates receives more citations (11.1 vs 8.2 citations per article). Productivity in terms of Scopus articles is highest in Natural sciences, mathematics and statistics, Engineering, manufacturing and construction, Information and Communication Technologies, and Health and welfare, and lowest in Arts and humanities, Business, administration and law, Education, and Social sciences, journalism and information. It should be noted, however, that the number of works declared in the compendium is very similar across fields (between 3.3 and 4.3 per thesis, excluding Services), so the differences reflect Scopus coverage and the publication channels typical of each field more than lower productivity among candidates. This was to be expected, since in the latter fields, in line with usual practice, the output linked to the thesis need not appear in internationally indexed outlets. However, the absence of a requirement to publish in international outlets may call into question the external certification of these theses, as they are not subject to the scrutiny of the most relevant journals.

Another finding that casts some doubt on the soundness of some theses by publication is that a substantial part of the candidates' output has been channelled through journals of so-called *grey* publishers. MDPI, the most prominent publisher in our study, accounts for 23.2% of the output, and four in ten candidates published at least one article with it. In fields such as Health and welfare, Engineering, manufacturing and construction, and Agriculture, forestry, fisheries and veterinary, more than half of the candidates published at least one work with this publisher. Moreover, 22 candidates (7.0% of authors with at least one Scopus article) published all their output exclusively in MDPI journals. This publisher has been widely criticised for questionable editorial practices, such as lax peer review (Oviedo-García, 2021), an aggressive special issue policy (Mills et al., 2025) and very short review times (Repiso et al., 2021), all of which may compromise the quality of the work. Nevertheless, some of these features may be strong reasons for candidates or their supervisors to choose these journals over those of other publishers.

In this study, we found that in 121 works linked to the compendiums the candidate did not hold a leading position in the byline (14.0% for men vs 9.2% for women), which may indicate a minor role in the research. This finding is striking because, by its nature, the main contribution to an article linked to a doctoral thesis should have been made by the author of the thesis, not by one of the other authors of the article. It also connects with the objections raised by some examiners, who feared that the publications were the work of the supervisor and the team rather than the student (Sharmini, 2016). Although scientific convention usually holds that the first author made the most relevant contribution to a work, we cannot rule out that other hierarchies or arrangements, departing from this customary agreement, have been established within the research group. Specifically, in our analysis 17 authors have no Scopus work as first or sole author (5.4% of theses with at least one Scopus article). We therefore consider that authorship criteria in theses by publication need to be clarified, since in Spain only about a third of the relevant policies state explicitly that the candidate must be the lead author of all the works linked to the thesis (Mason et al., 2025).

The analysis also detected other practices that may indicate risks to the integrity of the thesis, such as the inclusion of works published in journals that had previously been delisted from Scopus for poor editorial practices, works published in special issues guest-edited by the candidates themselves, and works still under review in journals but not yet accepted. Finally, we detected three duplicated articles in the sample, meaning that each of these articles was part of the compendiums of two different candidates (six theses in total). This is striking, since including an article in a thesis by publication usually means that none of the other co-authors can use it in a different thesis.

These results invite reflection on the incentives that sustain theses by publication. As Horta and Li (2023) warn, the centrality of publishing during the doctorate largely reflects how research staff are later recruited and evaluated, and a candidate who completes a thesis with a good number of publications in relevant journals is undoubtedly in a strong position in academic promotion processes. In Spain, the assessment of academic staff is moving towards more qualitative models, in line with the principles of CoARA (Coalition for Advancing Research Assessment) and DORA (San Francisco Declaration on Research Assessment), but publication in mainstream journals, such as those indexed in Scopus or the Journal Citation Reports (JCR), still plays an essential role in calls for academic posts. It has even been claimed that "a thesis through scientific articles offers, in principle, the greatest guarantee of the quality of a doctoral thesis, because an editorial committee impartially evaluates it, provided that these articles are accepted by scientific journals indexed in the JCR" (Serra-Aracil et al., 2022, p. 501). This logic, however, may reinforce a view of the thesis by publication centred on the number of articles, at the expense of other dimensions of doctoral training, such as research autonomy or the candidate's real contribution. Mertkan et al. (2022) note that repeated rejections fuel anxiety and insecurity among candidates, who come to see the most prestigious journals as out of reach. Within this mindset, and when obtaining the degree depends on publishing in an indexed journal, the alternative may in some cases be to turn to journals with questionable editorial practices.

*Limitations and future research*

This study has several limitations that may affect the results. The first is that it is not possible to obtain the full population of theses by publication defended in Spain, since no data source allows them to be identified automatically. Our study, based on the Teseo database, is therefore an approximation to this format through a sample, and the results could differ if the full population could be analysed. The limitations of the database itself, such as incomplete coverage or delays in indexing, may also affect the results. In addition, this study relies on manual processes, such as identifying the output of each compendium in the full text of the thesis and then searching for it in Scopus, so errors in these procedures are possible.

Moreover, Scopus coverage varies across fields and is lower in the social sciences and, above all, in the arts and humanities, where books, chapters and national journals are common, so article-based indicators may underrepresent the output of these fields. Another limitation is that citations were collected on a single date and were not normalised, so citation windows differ and comparisons between fields are not possible.

One methodological issue identified in this study is the lack of standardisation in how the compendium of articles is presented within the doctoral thesis. In some cases, the list is clearly set out in the opening pages of the thesis, while in others it appears in the Methodology section or in footnotes, which makes it very difficult to automate the detection of these works. We therefore call for the output linked to theses by publication to be identified in a more consistent and standardised way, so that thesis databases (such as Teseo or Dialnet) can include this information in a specific field and catalogue these theses as such, making them easier to retrieve and analyse. This would also make the output linked to embargoed theses available. Since the thesis by publication is now a common format in Spain because of its advantages for both candidates and supervisors (Codina, 2026), progress towards identifying these theses is increasingly necessary.

Finally, this study provides new empirical data on theses by publication in Spain and opens the way to deeper analyses of this format. Future research could compare the productivity of theses by publication with that of traditional theses, examine how this format affects professional careers, and explore which roles, according to standardised taxonomies, candidates take on in the research, which would give a better idea of whether they made substantial contributions even when they held intermediate positions in the byline. Last but not least, future studies should establish the percentage of theses by publication among all theses defended in Spain.

Doctoral theses emerged in the nineteenth century, within the German university, as tangible proof that there was a doctor: an autonomous researcher able to conceive, carry out and communicate original research on their own (Archbald, 2011). Theses by publication transfer that proof to a set of articles, usually co-authored, in which the candidate's individual contribution is hard to establish and quality is delegated to external editorial processes that do not always offer the guarantees assumed of them. The cases identified (publication with grey publishers or in delisted outlets, or the inclusion of works still under review) do not *per se* imply misconduct, but they weaken the very verification mechanisms that give the degree its value. Because theses by publication depend on the trust placed in these external review processes, further research is needed; otherwise, neither academic committees nor examination committees will have the tools or the overall view needed to detect possible irregularities, errors and shortcomings.

## Declaration of Competing Interest

This research did not receive any specific grant from funding agencies in the public, commercial, or not-for-profit sectors. The authors have no competing interests to declare that are relevant to the content of this article.

**Declaration of AI use**

Consensus was used to assist with the literature review. Byline position, the candidate's sex and the normalisation of publisher data were assigned with Claude (Sonnet 5). The translation of the manuscript from Spanish into English was carried out with Claude (Opus 5.5). All AI-assisted processes were reviewed and supervised by the authors.

**CRediT authorship contribution statement**

Álvaro Cabezas-Clavijo: Conceptualization, Data curation, Formal analysis, Investigation, Methodology, Supervision, Writing – original draft, Writing – review & editing.

Nazaret Barriga: Data curation, Investigation, Writing – review & editing.

Rafael Repiso-Caballero: Conceptualization, Methodology, Writing – original draft, Writing – review & editing.

Conception · Execution · Management · Leadership · Output

| AUTHORS | Conceptualization | Data curation | Formal analysis | Funding acquisition | Investigation | Methodology | Project administrati | Resources | Software | Supervision | Validation | Visualization | Writing – original dr | Writing – review & e | ROLES |
|---|---|---|---|---|---|---|---|---|---|---|---|---|---|---|---|
| ÁC Álvaro Cabezas-Clavijo #1 1st author | ✓ | ✓ | ✓ | | ✓ | ✓ | | | | ✓ | | | ✓ | ✓ | 8 |
| NB Nazaret Barriga #2 Author #2 | | ✓ | | | ✓ | | | | | | | | | ✓ | 3 |
| RR Rafael Repiso-Caballero #3 Last author | ✓ | | | | | ✓ | | | | | | | ✓ | ✓ | 4 |

*Generated with Credit Taxonomy Creator https://acabezasclavijo.github.io/Credit/*